\documentclass[fleqn,usenatbib,useAMS]{mnras}

\usepackage{graphicx}	% Including figure files
\usepackage[fleqn]{amsmath}	% Advanced maths commands
\usepackage{eqnarray}	% Equation array commands
\usepackage{amssymb}	% Extra maths symbols
\usepackage{multicol}        % Multi-column entries in tables
\usepackage{bm}		% Bold maths symbols, including upright Greek
\usepackage{pdflscape}	% Landscape pages

\usepackage[T1]{fontenc}
\usepackage{ae,aecompl}

\usepackage{newtxtext,newtxmath}
\title[Torques felt by solid accreting planets]{Torques felt by solid accreting planets}

\author[Zs. Reg\'aly]{Zs. Reg\'aly$^{1}$\thanks{Contact e-mail: \href{regaly@konkoly.hu}{regaly@konkoly.hu}}
\\
$^{1}$Konkoly Observatory, Research Centre for Astronomy and Earth Sciences, 1121, Budapest, Konkoly Thege Mikl\'os \'ut 15-17, Hungary}

\pubyear{2020}

\begin{document}
\label{firstpage}
\pagerange{\pageref{firstpage}--\pageref{lastpage}}
\maketitle

\begin{abstract}
The solid material of protoplanetary discs forms an asymmetric pattern around a low-mass planet ($M_\mathrm{p}\leq10\,M_\oplus$) due to the combined effect of dust-gas interaction and the gravitational attraction of the planet. Recently, it has been shown that although the total solid mass is negligible compared to that of gas in protoplanetary discs, a positive torque can be emerged by a certain size solid species. The torque magnitude can overcome that of gas which may result in outward planetary migration. In this study, we show that the accretion of solid species by the planet strengthens the magnitude of solid torque being either positive or negative. We run two-dimensional, high-resolution ($1.5\rm{K}\times 3\rm{K}$) global hydrodynamic simulations of an embedded low-mass planet in a protoplanetary disc. The solid material is handled as a pressureless fluid. Strong accretion of well-coupled solid species by a $M_\mathrm{p}\lesssim0.3\,M_\oplus$ protoplanet results in the formation of such a strongly asymmetric solid pattern close to the planet that the positive solid torque can overcome that of gas by two times. However, the accretion of solids in the pebble regime results in increased magnitude negative torque felt by protoplanets and strengthened positive torque for Earth-mass planets. For $M_\mathrm{p}\geq3\,M_\oplus$ planets the magnitude of the solid torque is positive, however, independent of the accretion strength investigated. We conclude that the migration of solid accreting planets can be substantially departed from the canonical type-I prediction.
\end{abstract}


%%%%%%%%%%%%%%%%%%%%%%%%%%%%%%%%%%%%%%%%%%%%%%%%%%

%%%%%%%%%%%%%%%%% BODY OF PAPER %%%%%%%%%%%%%%%%%%

\section{Introduction}
%------------------------------------------------------------------
Planets gravitationally interacting with their natal gaseous discs create two-armed spiral waves in the gas. The inner and outer spiral {\bf wakes} being stationary in the planetary reference frame exert positive and negative torques on the planet, respectively \citep{OglivieLubow2002}. In an isothermal disc assuming that the surface mass density distribution of gas follows a power law of $r^{-1}$ the total torque exerted by spiral waves is negative \citep{Ward1997}. As a result, the planet loses angular moment and migrates inwards in the so-called type~I regime. Since the migration speed is linearly proportional to the planet-to-star mass ratio \citep{GoldreichTremaine1980}, migration of growing planetary embryos (about at Mars mass) are negligibly slow \citep{Tanakaetal2002}. However, type~I migration of grown low-mass planets (about an Earth-mass, $M_\oplus$) can be fatal, i.e. they lost to the central star within a hundred thousand years, within about $6$\,Myr typical disc lifetime \citep{Hernandezetal2007}.  

Overall, no mechanism has been found that globally reduce the radial migration rate of low-mass planets \citep{MorbidelliRaymond2016}. Thus, to reproduce the statistics of observed planetary systems, planet synthesis models had to reduce the speed of type~I migration by about an order of magnitude (see, e.g.,  \citealp{IdaLin2004a,IdaLin2004b,Alibertetal2005,Migueletal2011}).

\citet{Massetetal2006} suggested that type~I migration trap may exist in places where the density gradients are positive, e.g., at disc inner edge, the boundaries of MRI active and inactive regions so-called dead zones or opacity transitions \citep{Masset2011}. However, two-dimensional hydrodynamic simulations showed that trapping of low-mass planets at viscosity transitions occur only if the dead zone edges are sharp enough to form a large-scale anticyclonic vortex \citet{Regalyetal2013}. Nonetheless, the existence of such planetary traps might explain the abundance of close-in super-Earths.

Another promising mechanism to slow down or even reverse the direction of type~I migration in non-isothermal discs is associated with sharp temperature gradients \citep{PaardekooperMellema2006,BaruteauMasset2008,PaardekooperPapaloizou2008}. Two and three-dimensional numerical simulations in non-isothermal radiative discs have shown that migration of low mass planets ($5-35\,M_\oplus$) can even reverse \citep{KleyCrida2008,Kleyetal2009,Legaetal2014}. Type~I torque depends on various physical parameters of the disc, such as viscosity and opacity, which are varying during the disc evolution.  In an evolving disc by applying non-isothermal type~I torque formulae (see, e.g., \citealp{MassetCasoli2010,Paardekooperetal2010,Paardekooperetsl2011}) the outward migration of low-mass planets is found to be confined in a radial range of the disc at about 3-9\,au \citep{HasegawaPudritz2011,HellaryNelson2012,Bitschetal2015}.

 Radiative effects of a solid accreting low-mass planet can significantly modify the torque felt by the planet \citep{Benitez-Llambay2015}. The so-called heating torque is positive and caused by the formation of asymmetric gaseous density lobes in the vicinity of the planet. The heat released by a vigorously accreting $3\,M_\oplus$ planet with a mass doubling time less than about $5000$ orbits results in migration reversal.  Studies considered neglect the solid torque exerted on low-mass planets. Since the dust-to-gas mass ratio is only about one percent in canonical protoplanetary discs (see, e.g., \citealp{WilliamsCieza2011}), it is plausible to assume that its gravitational effect dwarfed by that of gas.  However, the asymmetric distribution of solid matter in the vicinity of the planet might significantly modify the torque felt by that planet, similarly to the heating torque caused by the asymmetric gas distribution inside the planetary Hill sphere.  Indeed due to the combined effect of dust-gas interaction and the gravitational attraction of the planet, highly asymmetric distribution of solid material can form near low-mass planets. \citet{Benitez-Llambay2018} reveald that the spatial distribution of solid material can be extremely asymmetric such that it changes the speed or even alters the direction of type~I migration. 

To study the dynamics of solid material \citet{Benitez-Llambay2018} have taken into account the gravitational forces of the central star and the planet, and the aerodynamical drag force between gas and solid species. Planetary accretion, however, modifies the spatial distribution of solids by the removal of solid material from the vicinity of the planet. In this study, we present an investigation of how planetary accretion of solids modifies the spatial distribution of various size solid species and solid torque felt by low-mass planets.  We present hydrodynamical simulations of planet-disc interactions assuming that the solid content of the disc is a pressureless fluid.  We investigate the effect of planet mass in the range of $0.1-10\,M_\oplus$, size of solid species, the initial slope of gas profile,  accretion strength on the solid torque.

The paper is organised as follows: In Section~2, the applied hydrodynamical model and the initial conditions of gas and solids are presented. Section~3 presenting our results show a numerical convergence study and the effect of accretion strength on the total torque felt by low-mass planets. Section~4 discusses the torque profiles of different solid specie, the effect of solid accretion and the smoothing strength of planetary potential on the spatial distribution of solid material in the proximity of the planet.  The paper closes with our summary and conclusions in Section~5. In Appendix~A, test simulations on the dust solver is presented.

\section{Hydrodynamical model}
%------------------------------------------------------------------

\subsection{Governing equations}
%..................................................................

We investigate the effect of protoplanetary discs' solid material on the torque felt by an embedded low-mass planet using two-dimensional hydrodynamical simulations. The solid material is assumed to be a pressureless fluid whose dynamic is affected by the aerodynamic drag force arising due to the velocity difference between the gas and solid parcels. We use {\small GFARGO2} for this investigation, which is our extension to {\small GFARGO}, a GPU supported version of the {\small FARGO} code \citep{Masset2000}.

The dynamics of gas and solid fluids perturbed by the embedded planet is described by the following equations:
\begin{flalign}
&  \frac{\partial \Sigma_\mathrm{g}}{\partial t}+\nabla \cdot (\Sigma_\mathrm{g} {\bm{v}})=0, \label{eq:cont}\\
&  \frac{\partial \bm{v}}{\partial t}+(\bm{v} \cdot \nabla)\bm{v}=-\frac{1}{\Sigma_\mathrm{g}} \nabla P + \nabla \cdot \bm{T} -\nabla \Phi, \label{eq:NS}\\
&  \frac{\partial \Sigma_\mathrm{d}}{\partial t}+\nabla \cdot (\Sigma_\mathrm{d} \bm{u})=-\dot\Sigma_\mathrm{acc}, \label{eq:contd}\\
&   \frac{\partial \bm{u}}{\partial t}+(\bm{u} \cdot \nabla)\bm{u}=- \nabla\Phi+\bm{f}_\mathrm{drag}, \label{eq:NSd}
\end{flalign} 
where $\Sigma_\mathrm{g}$, $\Sigma_\mathrm{d}$ and $\bm{v}$, $\bm{u}$ are the surface mass density and velocity of gas and solid (being either dust particles or pebbles), respectively. The gas pressure is given by assuming a locally isothemral equation of state, for which case $P=c_\mathrm{s}^2\Sigma$, where $c_\mathrm{s}$ is the local sound speed. A flat disc approximation is used for which case the disc pressure scale-height is defined as $H=hR$, where the aspect ratio is set to $h=0.05$. 

{In Equation~(\ref{eq:NS}) $\bm{T}$ is the viscous stress tensor calculated as
\begin{equation}
    \bm{T} = \nu\left(\nabla \bm{v} + \nabla \bm{v}^T -\frac{2}{3}\nabla\cdot\bm{v}\bm{I}\right),
\end{equation}
where $\nu$ is the disc viscosity and $\bm{I}$ is the two-dimensional unit tensor, see details in \citet{Masset2002} for calculating $\bm{T}$ in cylindrical coordinate system. We use the $\alpha$ prescription of \citet{ShakuraSunyaev1973} for the disc viscosity in which case $\nu=\alpha c_\mathrm{s}H$. $\alpha=10^{-4}$ is applied which is in the order of the smallest effective viscosity presumably present in protoplanetary discs, e.g. arising due to vertical shear instability \citep{StollKley2016}. Note that the diffusion of solid species is neglected, which can only be done in a low-viscosity limit applied in this study (see, e.g., \citealp{YoudinLithwick2007}).

The gravitational potential of the system, $\Phi$, is calculated as
\begin{equation}
\Phi=-G\frac{M_*}{R}-G\frac{M_\mathrm{p}}{\sqrt{R^2+R_\mathrm{p}^2 - 2RR_\mathrm{p} \cos(\phi-\phi_\mathrm{p}) + (\epsilon H)^2}}+\Phi_\mathrm{ind},
\label{eq:phi_tot}
\end{equation} 
where $G$ is the gravitational constant, $R$, $\phi$ and $R_\mathrm{p}$, $\phi_\mathrm{p}$ are the radial and azimuthal coordinates of a given numerical grid cell and the planet, respectively. $M_*$ and $M_\mathrm{p}$ is the star and planet mass, respectively. The indirect potential, $\Phi_\mathrm{ind}$, is taken into account as a non-inertial frame, which co-rotates with the planet is used for the simulation. Note, however, that the effect of $\Phi_\mathrm{ind}$ presumably negligible in our case as the investigated planet-to-star mass ratio is small, $M_\mathrm{p}/M_*\leq10^{-5}$ and no significant large-scale asymmetries develop in the disc (see more details in  \citealp{RegalyVorobyov2017}). Both the gas and solid self-gravity are neglected. The planetary potential is smoothed by a factor of $\varepsilon H$ assuming $\varepsilon=0.6$, which is found to be appropriate for two-dimensional simulations \citep{Kleyetal2012,Mulleretal2012}. Note that the vertical scale-height of solids may differ from that of gas due to size-dependent speed of sedimentation to the disc midplane \citep{DullemondDominik2004}. However, applying a planetary potential that depends on the Stokes number of solids has no solid ground and requires further three-dimensional investigations. Thus, both solids and gas experience the same smoothing of planetary potential in our simulations. To shed light on the effect of smoothing strength on the solid torques we run additional simulations with $\varepsilon=0.3$ and $0.9$, see details in Section~\ref{sec:smoothing}.

In Equation\,(\ref{eq:NSd}) the drag force exerted by the gas on the solid is
\begin{equation}
\bm{f}_\mathrm{drag}=\frac{\bm{v}-\bm{u}}{\mathrm{St}/\Omega},
\end{equation}
where $\mathrm{St}$ is the Stokes number of the given solid species and $\Omega=(GM_*)^{1/2}R^{-3/2}$ is the local Keplerian angular velocity. Note that diffusion of solid material and the back-reaction of the solid onto the gas are neglected.

Equation~(\ref{eq:NSd}) is solved by a two-step numerical method. First, the source term, i.e., the right-hand side of Equation~(\ref{eq:NSd}) is calculated then it is followed by the conventional advection calculation. For the source term we use a fully implicit scheme (see details in \citealp{Stoyanovskayaetal2017,Stoyanovskayaetal2018}). The solid velocity is updated at every step as 
\begin{equation}
    \bm{u}^{n+1}=\bm{v}^{n}+\Delta t \bm{a}^{n}_\mathrm{g} - \frac{(\bm{v}^{n} - \bm{u} ^{n}) + \Delta t (\bm{a}^{n}_\mathrm{g}-\bm{a}^{n}_\mathrm{d})}{1+\Delta t/\tau_\mathrm{s}},
    \label{eq:solidsol-1}
\end{equation}
where $\tau_\mathrm{s}=\mathrm{St}/\Omega$ is the stopping time of the solid species, $\Delta t$ is the time-step applied, and $\bm{a}^{n}_\mathrm{g}$ and $\bm{a}^{n}_\mathrm{d}$ are the acceleration of gas and solid parcels due to the pressure gradient and gravitational forces:
\begin{flalign}
&   \bm{a}^{n}_\mathrm{g}=-\frac{1}{\Sigma_\mathrm{g}^{n}}\nabla P^{n}-\nabla\Phi^{n}, \\
&   \bm{a}^{n}_\mathrm{d}=-\nabla\Phi^{n}.
    \label{eq:solidsol-2}
\end{flalign}
The above expressions of solid velocity are equivalent to Equations\,(23)-(25) in \citet{Stoyanovskayaetal2018} by assuming that $\epsilon=0$, which corresponds to vanishing solid-to-gas mass ratio, i.e. $\Sigma_\mathrm{d}/\Sigma_\mathrm{g}\simeq0$. This assumption is valid as long as there is no significant solid enhancement. In this case the back-reaction of solid onto the gas (a term $-(\Sigma_\mathrm{d}/\Sigma_\mathrm{g})\bm{f}_\mathrm{drag}$ in the right-hand-side of equation (\ref{eq:NS})) can be neglected. With the above scheme, the effect of aerodynamic drag can be modelled for solid species that have stopping time that is much smaller than the time-step ($\Delta t\ll\tau_\mathrm{s}$). For pebbles that have large stopping time ($\Delta t\gg\tau_\mathrm{s}$), the method described above is only applicable if the orbit of solids are not crossing.

The accretion of solid material onto the planet represented by $\dot\Sigma_\mathrm{acc}$) in Equation~(\ref{eq:contd}) is modelled by a reduction of the solid density inside the planetary Hill sphere. We use a scheme similar to the prescription of gas accretion given in \citet{Kley1999}.  At each time step the solid density reduced by $1-\eta\Delta t$ inside the planet's Hill sphere with the radius of $R_\mathrm{Hill}=a(M_\mathrm{p}/3M_*)^{(1/3)}$, where $\eta$ is the strength of accretion.  With the above prescription the half-emptying time is $\log (2)/\eta$, i.e. about 2/3 orbital time for $\eta=1$.  Two accreting scenarios are investigated $\eta=1$ and $0.1$ referred to as strong and weak accreting scenarios.  The solid density reduction is done in two steps: first, one-third of the density is removed from the inner part of the Hill sphere ($\Delta R\leq0.75R_\mathrm{Hill}$), then in the second step, two-third of the density is removed from the innermost part of the Hill sphere ($\Delta R\leq0.45R_\mathrm{Hill}$). For simplicity, the planet mass is kept constant, i.e., the removed mass (and the momentum) is not added to the planet. As a result, the total solid mass and moment are strictly not conserved in the system. Note that strong accretion results in an accretion rate in the same order of magnitude applied by \citet{Benitez-Llambay2015} to model the effect of heating torque. The total mass accreted by the planet is negligible compared to the planet mass for all solid species.

\subsection{Initial and boundary conditions}
%..................................................................

We handled solid material with fix Stokes number throughout the simulation domain. We modelled multiple solid species in nine bins: $\mathrm{St}=0.01,\,0.1,1,\,2,\,3,\,4,\,5,$ and $30$. The back-reaction of the solids to the gas is neglected, thus multiple species of solids can be modelled in one simulation. According to \citet{SupulverLin2000} the Stokes number of a solid species having a physical size of $s$ can be given as
\begin{equation}
    \mathrm{St}=\frac{\rho_\mathrm{i}}{\rho_\mathrm{g}}\frac{s}{(1-f)\sqrt{8/\pi}c_\mathrm{s}+fC_\mathrm{D}/2 v_\mathrm{rel}}\Omega,
    \label{eq:St}
\end{equation}
where $\rho_\mathrm{i}$ is the volume density of the solid material. Parameter $f$ describes the transition between Epstein and Stokes regimes as $f=s/(s+\lambda)$. Assuming that protoplanetary discs are dominated by hydrogen molecule, the mean free path is $\lambda=3.34589 10^{-9}/\rho_\mathrm{g}$. In the Stokes regime $C_\mathrm{D}-0.44$ according to \citep{Whipple197}. By assuming a vertical pressure balance, the gas density in the disc midplane can be given as $\rho_\mathrm{g}=(1/2\pi)(\Sigma/H)$. The relative velocity between the gas and solid, $v_\mathrm{rel}$ can be given as
\begin{equation}
    v_\mathrm{rel}=\sqrt{\left( v_R-u_R\right)^2+\left(v_\phi-u_\phi \right)^2}.
\end{equation}
Taking into account Equations,(\ref{eq:vel1})-(\ref{eq:vel2}), the relative velocity between the gas and dust is $v_\mathrm{rel}\leq0.01$ in a Minimum Solar Mass Nebula (MMSN) model \citep{Hayashi1981}. The solid physical size as a function of the Stokes number, $s(\mathrm{St})$, is shown in Fig.\,\ref{fig:d-St} at three distances (0.1, 1, and 10\,au) from the star. The solid bins investigated roughly represent the following regimes in the planet forming region at about several astronomical units distance from the star: $\mathrm{St}0.01$ is the mm-sized dust regime, $\mathrm{St}=0.1-1$ is the cm-sized pebble regime, $\mathrm{St}=1-10$ is the m-sized boulder regime, and $\mathrm{St}>10$ is the asteroid regime.

\begin{figure}
	\includegraphics[width=\columnwidth]{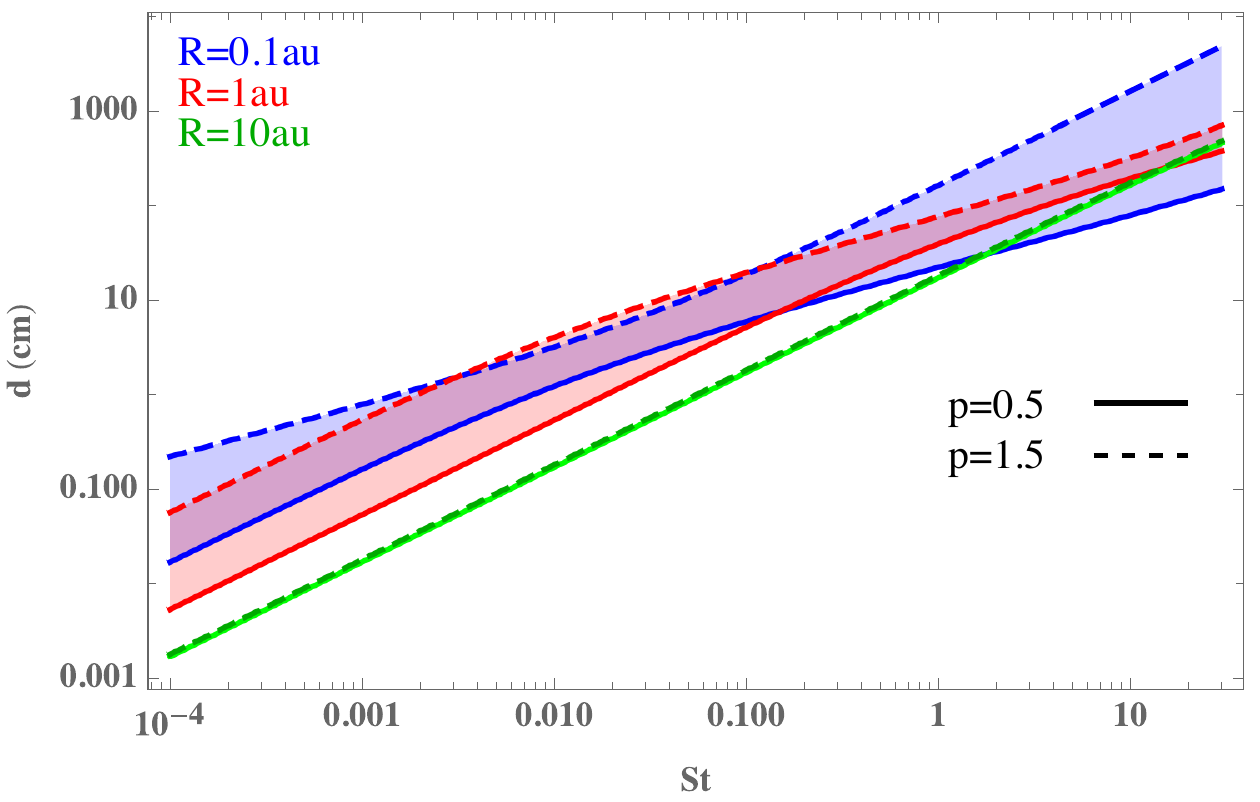}
    \caption{Solid size as a function of Stokes number at three distances: $R=0.1,\,1$ and $10$au. Solid and dashed lines represent same mass discs with initial density profile of $p=0.5$ and $p=1.5$, respectively. Shaded regions correspond for regions where the solutions are for $0.5 < p< 1.5.$}
    \label{fig:d-St}
\end{figure}

The mass of planets in our models are in the range of $q=3\times10^{-7}-3\times10^{-5}$, which correspond to $0.1M_\oplus-10M_\oplus$. We investigate five models in which planet mass in logarithmic bins is $M_\mathrm{p}=0.1,\,0.3,\,1,\,3$ and $10\,M_\oplus$. The orbital distance of the planet is set to unity, and the planet is kept on a circular orbit, i.e., no migration is allowed.

Initially $\Sigma_\mathrm{g}=\Sigma_0R^{-p}$ and $\Sigma_\mathrm{d}=\epsilon\Sigma_\mathrm{g}$ assuming $\Sigma_0=6.45\times10^{-4}$ and solid-to-gas mass ratio $\xi=10^{-2}$. We investigate three different initial gas and solid density profile steepness: $p=0.5,\,1.0$ and $1.5$ assuming fixed $\Sigma_0$. Emphasise that the absolute value of $\Sigma_\mathrm{g}$ does not affect our results for the following reasons: 1) disc self-gravity is neglected; 2) the amount of solid contained by the disc is scaled with the gas mass; 3) the dust back-reaction is neglected. 

The initial velocity components of gas ($v_R$ and $v_\phi$) are defined as 
\begin{flalign}
&   v_R=-3\alpha h^2(1-p)\Omega R, \label{eq:vel1}\\
&   v_\phi=\sqrt{1-h^2(1+p)}\Omega R.\label{eq:vel2}
\end{flalign}
The above equations satisfy the
steady-state solution to the viscous evolution of the surface mass density in $\alpha$ discs for $p=0.5$ and $p=1$. The initial velocity components of solid material ($u_R$ and $u_\phi$) are given by the analytic solution of an unperturbed disc, which reads 
\begin{flalign}
&   u_R=\frac{v_\mathrm{R}\mathrm{St}^{-1}-h^2 (1+p)}{\mathrm{St} + \mathrm{St}^{-1}}\Omega R, \label{eq:vel3}\\
&   u_\phi=\sqrt{1-h^2(1+p)}\Omega R - \frac{1}{2}u_R\mathrm{St}, \label{eq:vel4}
\end{flalign}
(see, e.g.,  \citealp{Nakagawaetal1986,TakeuchiLin2002}).

We use a computational domain whose extension is $0.48\leq R\leq2.08$ in code units, which contains 1536 logarithmically distributed radial and 3072 equidistant azimuthal grid cells. With these settings, the numerical resolution is about $0.02H$ at all distances. We confirmed that simulations with that resolution are in the numerically convergent regime (see Section~\ref{sec:numres}). 

At the inner and outer boundaries a wave damping boundary condition is applied for the gas (see, e.g., \citealp{deValBorroetal2006}). An open inner boundary condition is applied for the solid material. However, solids are replenishing at the outer boundary due to the applied damping boundary condition there. As a result, the solid disc is not depleted.

We assume 1\,au for the unit of length, and the stellar mass for the unit of mass  and the gravitational constant, G, is defined 1. With these assumptions, the orbital period is 2$\pi$ at 1\,au.

\section{Results}
%------------------------------------------------------------------

\subsection{Resolution required for numerical convergency}
%..................................................................
\label{sec:numres}

\begin{figure}
	\includegraphics[width=\columnwidth]{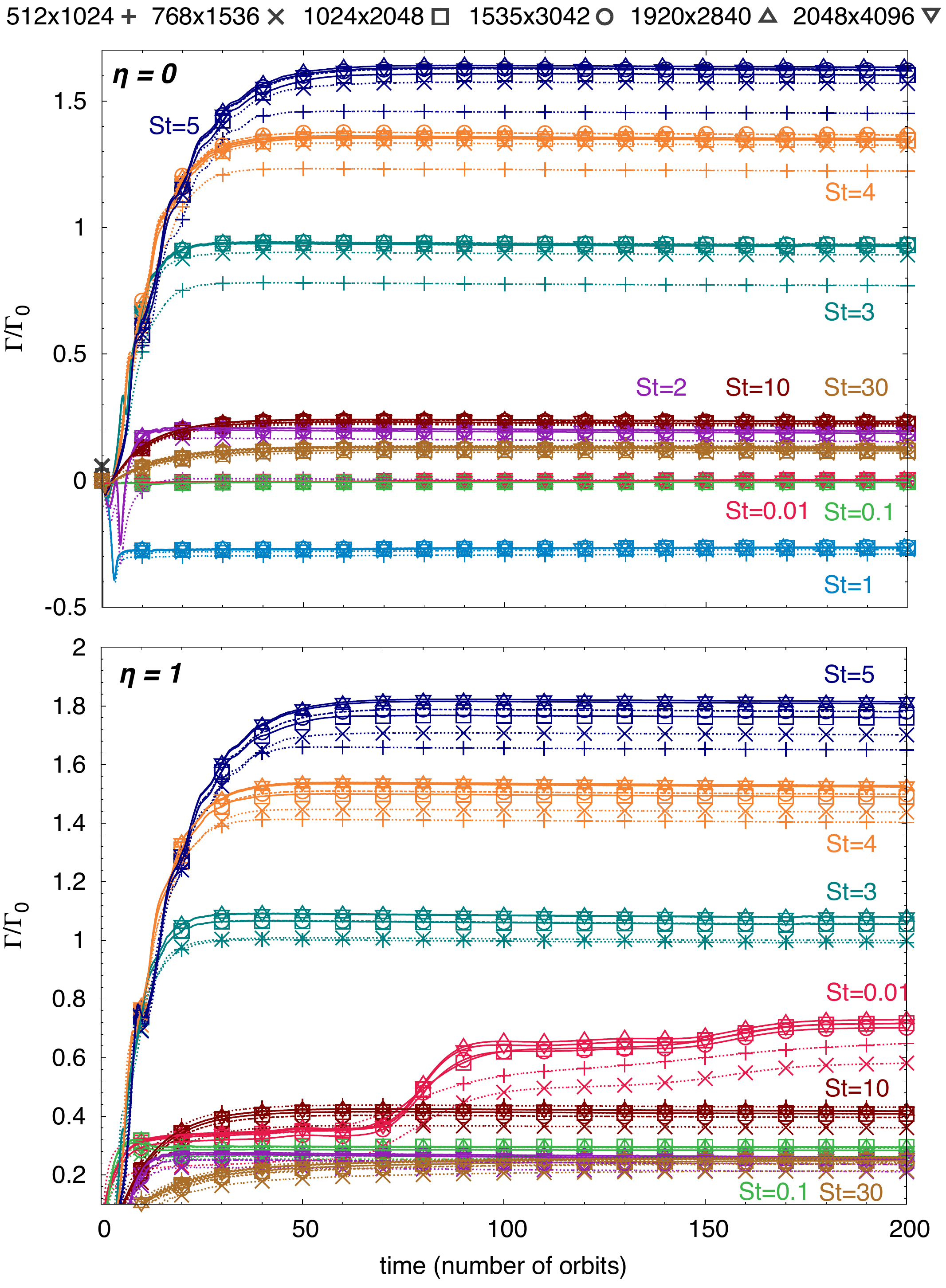}
    \caption{Numerical convergence test using six different numerical resolutions indicated in the top of the panel with different symbols. The figure shows the evolution of solid torques, $\Gamma/\Gamma_0$, felt by $1\,M_\oplus$ planet of solid species indicated in different colours. Upper and lower panels show the non-accreting and strong accreting models, respectively.}
    \label{fig:numres}
\end{figure}

First, we test that the hydrodynamical solutions are in the numerically convergent regime. We run non-accreting and strong accreting models assuming {\bf $q=3\times10^{-6}$} with different numerical resolutions: $N_R\times N_\phi = [512\times1024\,,768\times1536\,,1024\times2048\,,1536\times3072\,,1920\times3820\,,2048\times4096]$. 

Both the solid and gas torques are calculated as
\begin{equation}
    \Gamma=\Gamma_0\sum_{i,j=1}^{Nr,N\phi}\left(x_\mathrm{p}\frac{\Sigma_{i,j}}{\Delta R_{i,j}^3}-y_\mathrm{p}\frac{\Sigma_{i,j}}{\Delta R_{i,j}^3}\right ),
\label{eq:torques_beg}
\end{equation}
where
\begin{flalign}
&  \Delta R_{i,j}=\sqrt{\Delta x_{i,j}^2+\Delta y_{i,j}^2},\\
&  \Delta x_{i,j}=R_{i,j}\cos(\phi_{i,j})-x_\mathrm{p},\\
&  \Delta y_{i,j}=R_{i,j}\sin(\phi_{i,j})-y_\mathrm{p}, \label{eq:torques_end}
\end{flalign}
where $R_{i,j}$ and $\phi_{i,j}$ are the cylindrical coordinates of cell $i,j$, $\Sigma_{i,j}$ is the surface mass density of gas or the given solid species inside that cell. $x_\mathrm{p}$, $y_\mathrm{p}$ are the Cartesian coordinate of the planet. To compare torques felt by different mass planets it is useful to normalise the torques. The change in the semi-major axis, $a$, of a planet due to the torque, $\Gamma$, exerted by the gas and solid species can be given as $da/dt=2\Gamma/(qa\Omega(a))$. Thus, $\Gamma_0=2/(qa\Omega(a))$ is used as a normalisation factor throughout this paper. Since $a=1$ in our models, the normalisation constant is $\Gamma_0=2/q$.

The evolution of normalised torque of solid species, $\Gamma_d/\Gamma_0$, is shown in Figure~\ref{fig:numres}. The torque exerted by the gas saturates at $\Gamma_\mathrm{g}/\Gamma_0\simeq-0.8$ within 10 orbits. The gas torque is found to be independent of the numerical resolution therefore it is not shown in the Figure. However, the saturation value of the solid torque depends on the Stokes number and the numerical resolution too. In non-accreting models,  torque magnitudes are significant and sensitive to the numerical resolution for $1\leq \mathrm{St}\leq10$.  However,  in the strong accreting models solid torques become sensitive to the numerical resolution for the well-coupled species ($\mathrm{St}\leq0.01$ ) too. Although the effect of solid accretion (see details in the next section) can be identified in model that uses the lowest numerical resolution, a relatively high numerical resolution is required for proper torque values.

We found that the magnitude of solid torques is converged in models that use numerical resolution above $1024\times2048$. Therefore, a relatively fine resolution $N_R\times N_\phi=1536\times3072$ is used in this study. With this numerical resolution the planetary Hill sphere is resolved by about 72 cells for the smallest mass planet modelled.

\subsection{Effect of solid accretion}
%.................................................................

%\begin{figure}
%	\includegraphics[width=\columnwidth]{torque-evol}
%    \caption{Normalised gas and solid torques exerted on the $1M_\oplus$ planet as a function of time assuming different solid spices indicated with colours. Top to bottom panels show three set of models assuming $p=0.5,\,1.0$ and $1.5$, respectively. The small, medium and large symbols represent models assuming accretion efficiency of solid $\eta=0\%,\,10\%$ and $100\%$, respectively. The black line shows $-\Gamma_\mathrm{g}/\Gamma_0$, i.e. the magnitude of normalised gas torque. Those solid species for which case $\Gamma_\mathrm{d}/\Gamma_0>-\Gamma_\mathrm{g}/\Gamma_0$ exert net positive torque on the planet.}
%    \label{fig:gamma_mod}
%\end{figure}

\begin{figure}
	\includegraphics[width=\columnwidth]{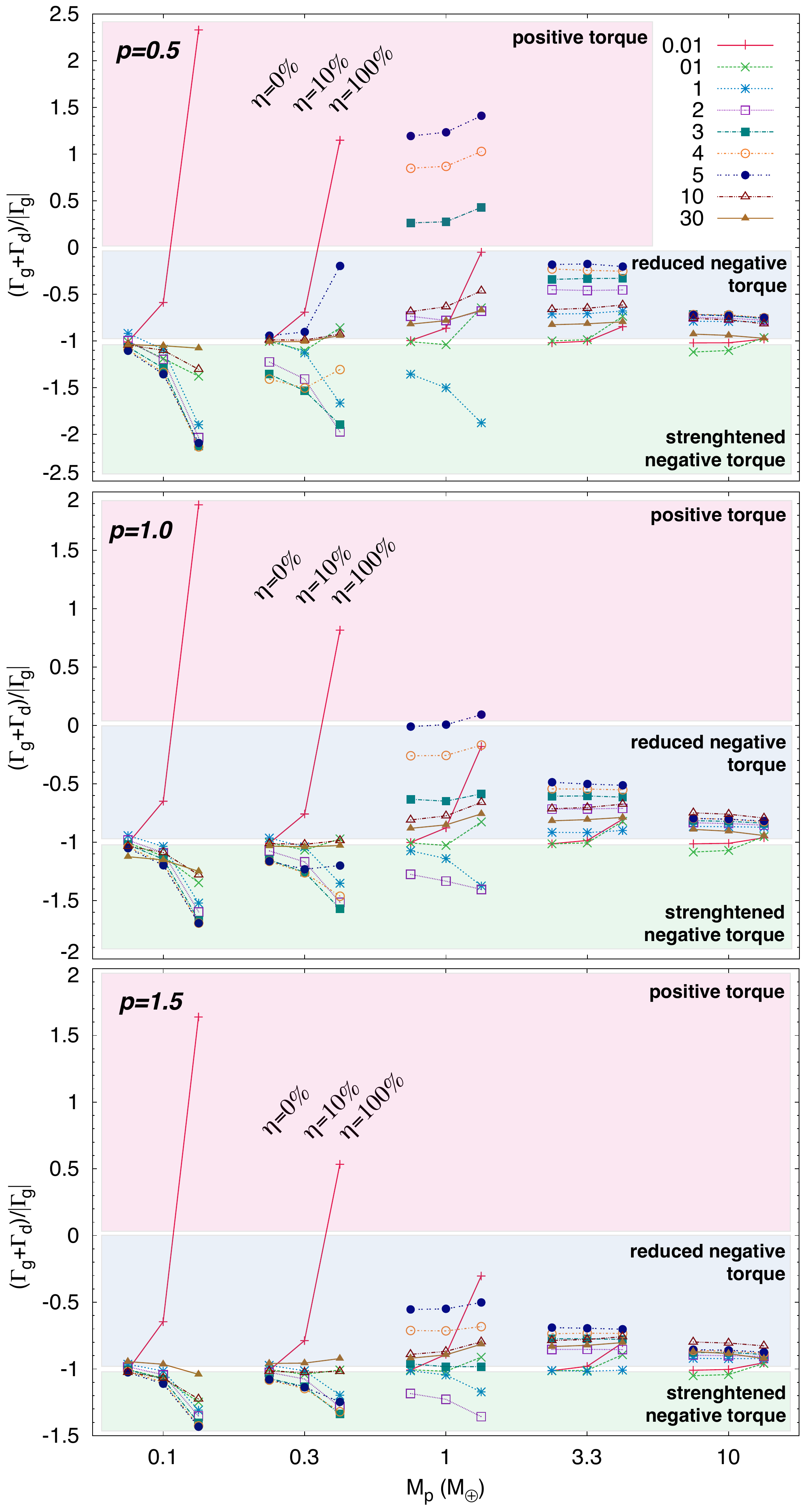}
    \caption{Effect of accretion strength, $\eta$, on the total torque normalised with absolute torque value of gas, $(\Gamma_\mathrm{g}+\Gamma_\mathrm{d})/|\Gamma_\mathrm{g}|$, felt by a planet with a mass in the range of $0.1\,M_\oplus\leq M_\mathrm{p}\leq10\,M_\oplus$ for different solid species indicated on the legend. Top to bottom panels show three set of models assuming $p=0.5,\,1.0$ and $1.5$, respectively. For each planet mass three accretion strength are investigated: $\eta=0,\,0.1$ and $1$. Green, blue and red shaded regions correspond to strengthened negative, reduced negative, and positive torques, respectively. Emphasise that the range of Y axis is decreasing with increasing $p$.}
    \label{fig:gamma_end}
\end{figure}

In this section, the effect of solid accretion on the torque felt by the planet is investigated. Fig.\,\ref{fig:gamma_end} shows the total torques measured at the end of simulation, after 200 orbits of the planet. Torques are normalised with the absolute value of the gas torque, $(\Gamma_\mathrm{g}+\Gamma_\mathrm{d})/|\Gamma_\mathrm{g}|$.  Five different planet mass ($0.1,\,0.3,\,1,\,3$ and $10\,M_\oplus$)  are investigated with three different strength of accretion, $\eta=0,\,0.1$ and $1$. The three panels show our results assuming three different steepness ($p=0.5,\,1.0$ and $1.5$) of the initial gas profile. 

Based on the value of the normalised total torque three different regimes can be defined:  $(\Gamma_\mathrm{g}+\Gamma_\mathrm{d})/|\Gamma_\mathrm{g}|>0$ shaded with red, $-1<(\Gamma_\mathrm{g}+\Gamma_\mathrm{d})/|\Gamma_\mathrm{g}|<0$ shaded with blue, and  $(\Gamma_\mathrm{g}+\Gamma_\mathrm{d})/|\Gamma_\mathrm{g}|<-1$ shaded with green colours. %By assuming that our simplifications (e.g. planet kept on fixed circular orbit) do not affect planetary migration significantly, red, blue and green shaded regions would correspond to outward, slowed down inward and accelerated inward migration regimes, respectively.
It is appreciable that solid accretion increases the magnitude of solid torque (for both positive or negative torque values) independent of $p$.

Let's investigate the model where $M_\mathrm{p}=1\,M_\oplus$ and $p=0.5$ , see upper panel of Fig.\,\ref{fig:gamma_end}. In all cases, except $\mathrm{St}=3$, the solid torques are positive. For $3\leq\mathrm{St}\leq5$, the solid torque is positive and its magnitude exceeds that of gas, which results in positive total torque. For $\mathrm{St}=3$, the solid torque is negative, therefore the planet feels stringer negative total torque with increasing accretion strength. For $\mathrm{St}>5$ and $\mathrm{St}<0.1$, the magnitude of solid torques are positive and has a small amplitude (except cases of strong accretion), therefore solid material slightly decreases the magnitude of negative total torque. 

\begin{figure*}
\includegraphics[width=2\columnwidth]{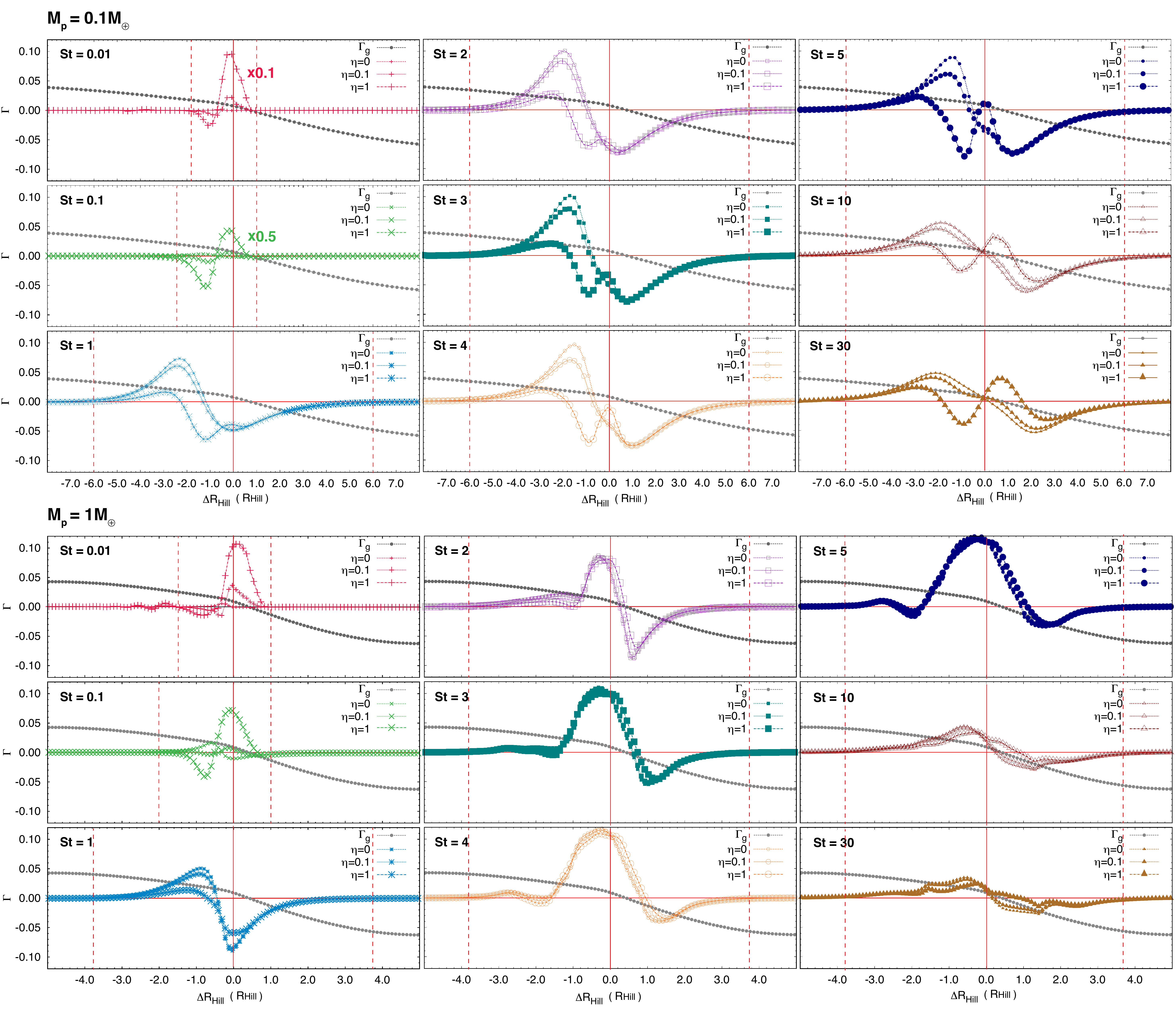}
    \caption{Azimuthally averaged radial torque profiles of dust species in the vicinity of a $0.1M_\oplus$ (upper panels) $1M_\oplus$ planet (lower panels). The gas torque profile is also shown with black symbols. The radial distance is measured in units of the Hill sphere of the given planet. Note that the spatial region displayed for the upper and lower panels are the same. Emphasize that torques of $\mathrm{St}=0.01$ and $0.1$ species are multiplied by 0.1 and 0.5 in case of $M_\mathrm{o}=0.1\,M_\oplus$ planet. Vertical red dashed lines mark the non-zero solid torque regions.}
    \label{fig:Torque-prof}
\end{figure*}

For $M_\mathrm{p}\geq3\,M_\oplus$ planets the total torque amplitude is nearly independent of $\eta$, i.e., the effect of solid accretion is negligible. However, for smaller planet mass ($M_\mathrm{p}\leq1\,M_\oplus$), solid accretion can have a severe effect on planetary torque: planet can feel two times larger magnitude torque compared to that of conventional type~I regime. For low-mass planet ($M_\mathrm{p}\leq0.3\,M_\oplus$) positive total torque can be observed if it strongly accretes well-coupled $\mathrm{St}=0.01$ solid species.

Two additional panels of Fig.\,\ref{fig:gamma_end} show models in which different slopes for the gas density profile, $p=1$ and $1.5$ are used. The steeper the initial density profile, the more negative the normalised total torque. Emphasise that $M_\mathrm{p}\leq0.3\,M_\oplus$ planets that strongly accrete $\mathrm{St}\leq0.1$ solids experience positive torque. Another effect of the steep initial gas profile is the moderate growth of negative torque amplitude (about 1.5 times) due to solid accretion.

In summary, solid accretion generally increases the magnitude (either positive or negative) of the total torque felt by the planet. The total torque can be positive if the accretion of the well-coupled solid species is strong. For about an Earth-mass planet, the pebble accretion ($3\leq\mathrm{St}\leq5$), independent of accretion strength, can also lead to positive torque (for $p=0.5$) or strongly reduced (for $p=1$ or $p=1.5$) negative torque. The effect of accretion strength on the total torque felt by the planet weakens with steeper radial density profile of gas.

\section{Discussion}
%------------------------------------------------------------------

\begin{figure}
	\includegraphics[width=1\columnwidth]{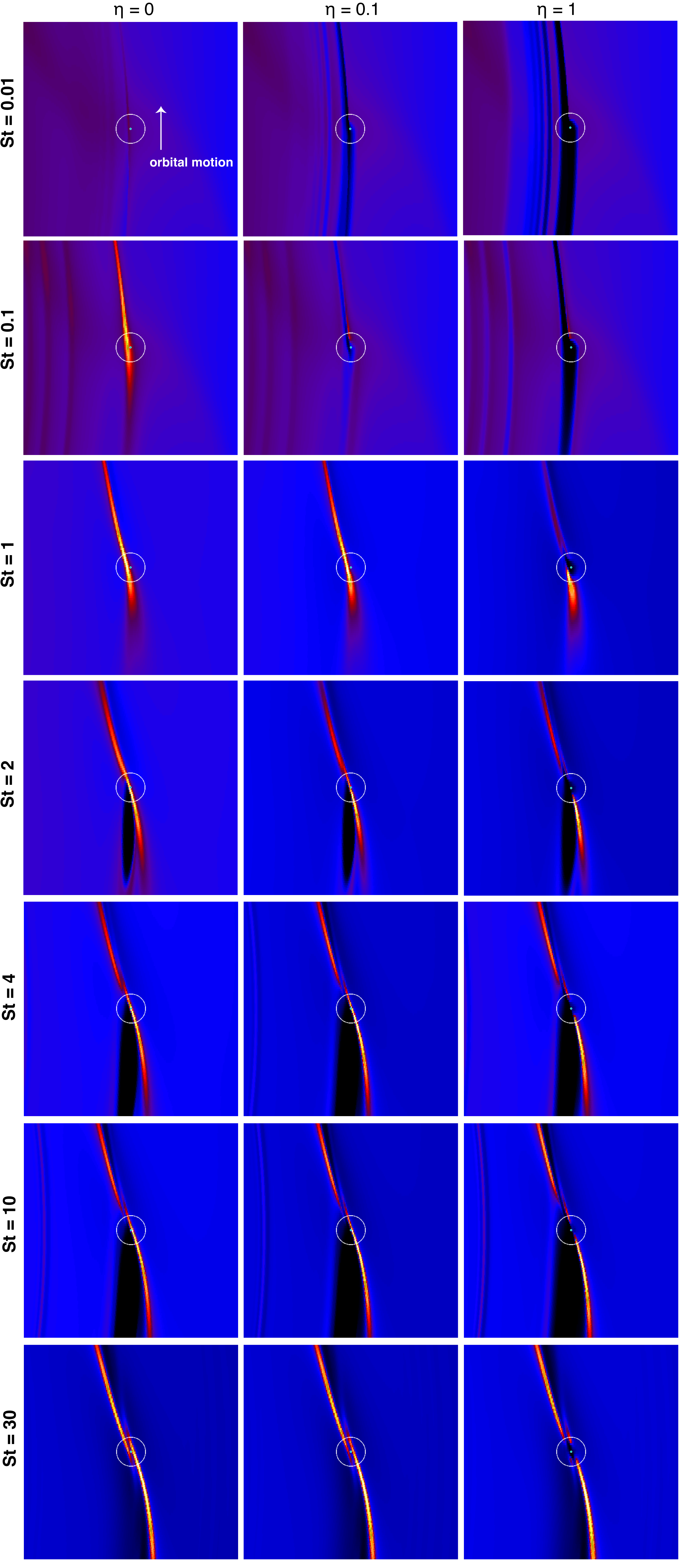}
   \caption{Comparison of solid density distribution of species having $\mathrm{St}=0.01,\,0.1,\,1,\,2,\,4,\,10$, and $30$ in the three accretion regimes for $1\,M_\oplus$ planet assuming $p=0.5$. White circles show the planetary Hill sphere. Black-blue-red colours represent growing solids density. The orbital motion of the planet is indicated on the first panel.}
    \label{fig:soliddens}
\end{figure}

\subsection{Torque profiles}
%..................................................................

First, let's investigate the radial profiles of torques exerted by the gas and different solid species.  Fig.\,\ref{fig:Torque-prof} shows the azimuthally averaged radial torque profiles in the three accretion regimes as a function of the radial distance from the planet measured in units of the radius of planetary Hill sphere. Two models  assuming $0.1\,M_\oplus$ and $1\,M_\oplus$ planets are shown on the top and bottom panels of Fig.\,\ref{fig:Torque-prof}, respectively. Note that the spatial region covered by the plots are the same for all panels.

The gas torque arises from a region that has an extent of about $\pm10R_\mathrm{Hill}$ and it is independent on the applied accretion  strength. The latter statement is valid as long as the solid back-reaction and the self-gravity of the disc are neglected. The inner and outer gas disc exerts exclusively positive and negative torques, respectively. Since $\Gamma_\mathrm{g}$ profile is only slightly asymmetric a small non-negligible negative gas torque is exerted on the planet as the theory predicts (see, e.g., \citealp{Ward1997}). 

On the contrary, solid torque profiles are highly non-symmetric and their shapes are sensitive to the Stokes number of the given species and the accretion strength too. Torques of solid species generally vanish beyond $\pm6R_\mathrm{Hill}$. Independent of the planet mass, the solid torques of the well-coupled solid species ($\mathrm{St}\leq0.1$) arising from a region of $\left[-3\Delta R_\mathrm{Hill},1\Delta R_\mathrm{Hill}\right]$. This means that the physical distance from within the well-coupled solid species can exert torque on the planet is roughly bound to the planetary Hill sphere, namely it depends on the planet mass. It is also appreciable that the torque profile of $\mathrm{St}=0.01$ becomes positive as the strength of solid accretion increases. This explains why the torque exerted by the well-coupled solid species can even overcome that of gas. 

Solid torques for $\mathrm{St}\gg1$ are non-vanishing up to a distance of $\pm 6 \Delta R_\mathrm{Hill}$ and $\pm 3 \Delta R_\mathrm{Hill}$ in the case of $0.1\,M_\oplus$ and $1\,M_\oplus$ planets, respectively. Thus, the extent of the influencing region of the less coupled solid species is generally independent of the planet mass. 

The torque profile can be significantly affected by the accretion of less coupled solid species, $\mathrm{ST}>1$. The general trend is that the positive part of the torque profile diminishes with increasing $\eta$. As a result, solids tend to increase the magnitude of negative torque. This phenomenon is prominent for $M_\mathrm{p}=1\,M_\oplus$ planets up to $\mathrm{St}=10$, while suppressed for larger mass planets. Consequently, the accretion profiles are insensitive to the accretion of $\mathrm{St}>1$ solids for larger mass planets ($M_\mathrm{p}\geq1\,M_\oplus$), which  explains the decreasing amplitude of solid torques  for larger mass planets (see Fig.\,\ref{fig:gamma_end}).  Note, however, that the torque profile of pebbles ($1<\mathrm{St}\leq5$) is mainly positive for larger mass planets.

\subsection{Distribution of solids in the vicinity of planet}
%-----------------------------------------------------------------------

Based on the fact that the density of solids determines the torque magnitude, see Equations\,(\ref{eq:torques_beg}), it is worth investigating its spatial distribution.  Fig.\,\ref{fig:soliddens} shows the solid density distribution of all investigated solid species in the vicinity of $1\,M_\oplus$ planet in $p=0.5$ disc at the end of the simulation, when a quasi-steady state has been developed. It can be seen that solids form a spiral-like overdense pattern. Emphasise that these patterns do not coincide with the gas spiral, except for the well-coupled solid species, $\mathrm{St}=0.01$. The opening angle of spiral patterns is increasing with the Stokes number of solid species.

Another prominent feature of the solid distribution is the development of a strong density depletion behind the planet (concerning planetary orbital motion) for the non-coupled solid species ($\mathrm{St}>1$). The size of the solid depleted region grows with both the Stokes number and $\eta$. Emphasise that the solid depleted region can be developed for the well-coupled dust species ($\mathrm{St}\leq0.1$) too in accreting models (see panels $\eta=1$, $\mathrm{St}=0.01$ and $\mathrm{St}=0.1$ of Fig.\,\ref{fig:soliddens}). Due to the formation of these overdense and depleted solid patterns, the solid distribution can become highly non-symmetric. As a result, the disc region beyond and behind the planet can exert different magnitude and sign torques. 

Now, let's investigate the change in the spatial distributions of solid species due to their accretion by calculating the value of $[\Sigma_d(\eta=1)-\Sigma_d(\eta=0)]/\Sigma_d(\eta=0)$ in the vicinity of the planet. In Fig.\,\ref{fig:comparison} red colour corresponds to regions where the distribution of solids is unaltered by the solid accretion, while blue coloured regions reveal strong depletion due to solid accretion. In the strong accreting model, the solid depleted region formed beyond the planet is narrower than behind the planet for the well-coupled solid species, $\mathrm{St}\leq0.1$. Therefore, in case of strong accretion the positive torque arising from the disc region beyond the planet overcomes the negative torque exerted by the disc region behind the planet. The bright vertical structures inside the planetary orbit appear due to a solid depleted in-spiralling pattern present in the strong accreting model. For $\mathrm{St}=1$, only a small depletion developed very close to the planet as a result of solid accretion, which causes a negative net solid torque (see panel $\mathrm{St}=1$ on Fig.\ref{fig:Torque-prof}). Solid species with $2\leq\mathrm{St}\leq5$ generally form a depleted region behind the planet which is wider in the strong accreting regime. As a result, a positive solid torque appears, whose magnitude increases with accretion strength. For the decoupled solid species ($\mathrm{St}\geq10$) the accretion results in symmetric solid removal inside the planetary Hill sphere. Hence the torque profiles are damped and do not change significantly due to accretion (see panels $\mathrm{St}=10$ and $\mathrm{St}=30$ on Fig.\ref{fig:Torque-prof})). 

By examining the solid distribution in models where the initial gas density profile is steeper, we found no significant departure to that of $p=0.5$ model. As a result, the solid torques have a similar dependence on $\eta$ and magnitudes independent of $p$.  However, due to the fact that the gas torque increases with $p$ ($\Gamma_\mathrm{g}/\Gamma_0=0.75,\,0.95$, and $1.18$ for $p=0.5,\,1.0$, and $1.5$, respectively). solids exert lowered torque on the planet in models that have steeper gas density slope.

\begin{figure}
	\includegraphics[width=1\columnwidth]{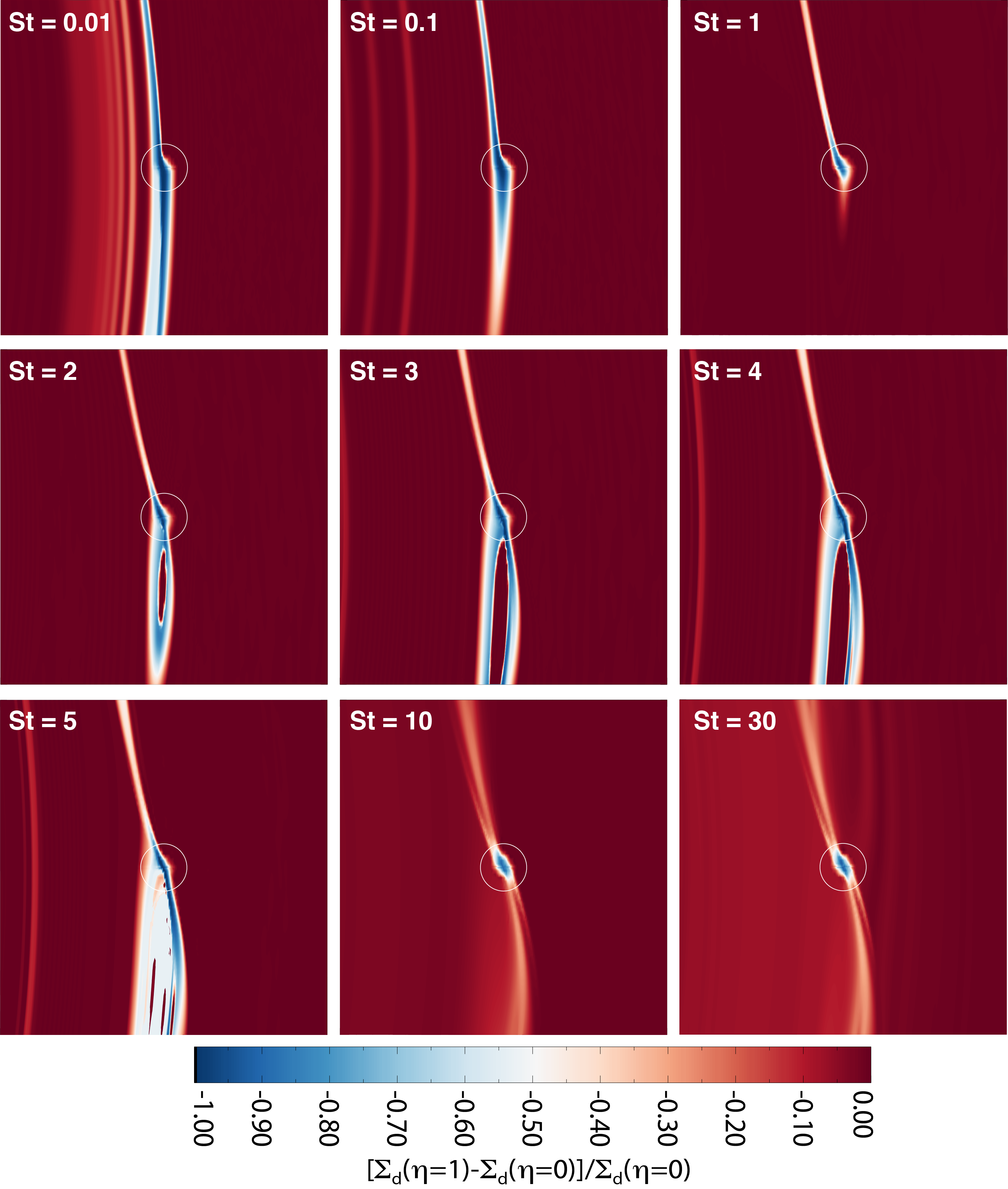}
   \caption{Comparison of solid density distributions in the vicinity of $1\,M_\oplus$ planet in non-accreting and strong accreting models. Comparison is calculated as  $[\Sigma_\mathrm{d}(\eta=1)-\Sigma_\mathrm{d}(\eta=0)]/\Sigma_\mathrm{d}(\eta=0)$. Regions where the two models are the same appear in red colour. In the strong accreting models, blue regions are completely empty of solid species.}
    \label{fig:comparison}
\end{figure}

\subsection{Comparison to previous work}
%..................................................................

To date, only one work addressed the solid torque felt by low-mass ($0.3\,M_\oplus\leq M_\mathrm{p}\leq10\,M_\oplus$) non-accreting plants. \citet{Benitez-Llambay2018} investigated the effect of solids in a disc with $p=0.5$ and similar spatial extensions. They used a somewhat more viscous disc, $\alpha=3\times10^{-3}$, however, the dust diffusion was also neglected in their simulations. They stated that the majority of simulations reach steady-state, which was confirmed by our runs. Note, however, that to reach steady-state requires 200 orbits for the well-coupled solid species, see Fig.\,\ref{fig:numres}.

According to Figure\,2 of \citet{Benitez-Llambay2018}  planets generally feels positive total torque for $\mathrm{St}\gtrsim0.1$ solid species as $\log(\Gamma_\mathrm{d}/|\Gamma_\mathrm{g}|)\geq0$ for those cases. Some exceptions, however, can be identified for $0.1<\mathrm{St}<1$ and $M_\mathrm{p}\leq1\,M_\oplus$ cases. These findings are practically confirmed by our $M_\mathrm{p}\geq0.3\,M_\oplus$ simulations as the normalised total torques are in the blue or red shaded region for all non-accreting models upper panel of Fig.\,\ref{fig:gamma_end}. Note, however, that for our $M_\mathrm{p}=0.3\,M_\oplus$ model, solids in the pebble regime ($2\leq\mathrm{St}\leq 4$) provide negative torque. The discrepancy can be attributed to the difference in torque calculation in the two studies or the difference in the applied magnitude of viscosity. The latter requires further study as diffusion of solids (especially with small Stokes number) should be taken into account in non-inviscid disc. 

\citet{Benitez-Llambay2018} completely neglect torques arising from the inner half of the planetary Hill sphere, while we take in to account the entire planetary Hill sphere. Note that the idea of torque cut-off was introduced by \citet{Cridaetal2009} to calculate disc torques exerted on massive planets to account their circumplanetary discs in non-self-gravitating models. \citet{Cridaetal2009} conclude that removing the half of the material of the planetary Hill sphere is appropriate. Since the circumplanetary disc does not form around low mass ($M_\mathrm{p}\lesssim30\,M_\oplus$) planets \citep{Massetetal2006}, the exclusion of material in the planetary Hill sphere has no physical argument.

\subsection{Effect of smoothing of planetary potential}
%..................................................................
\label{sec:smoothing}

\begin{figure}
	\includegraphics[width=1\columnwidth]{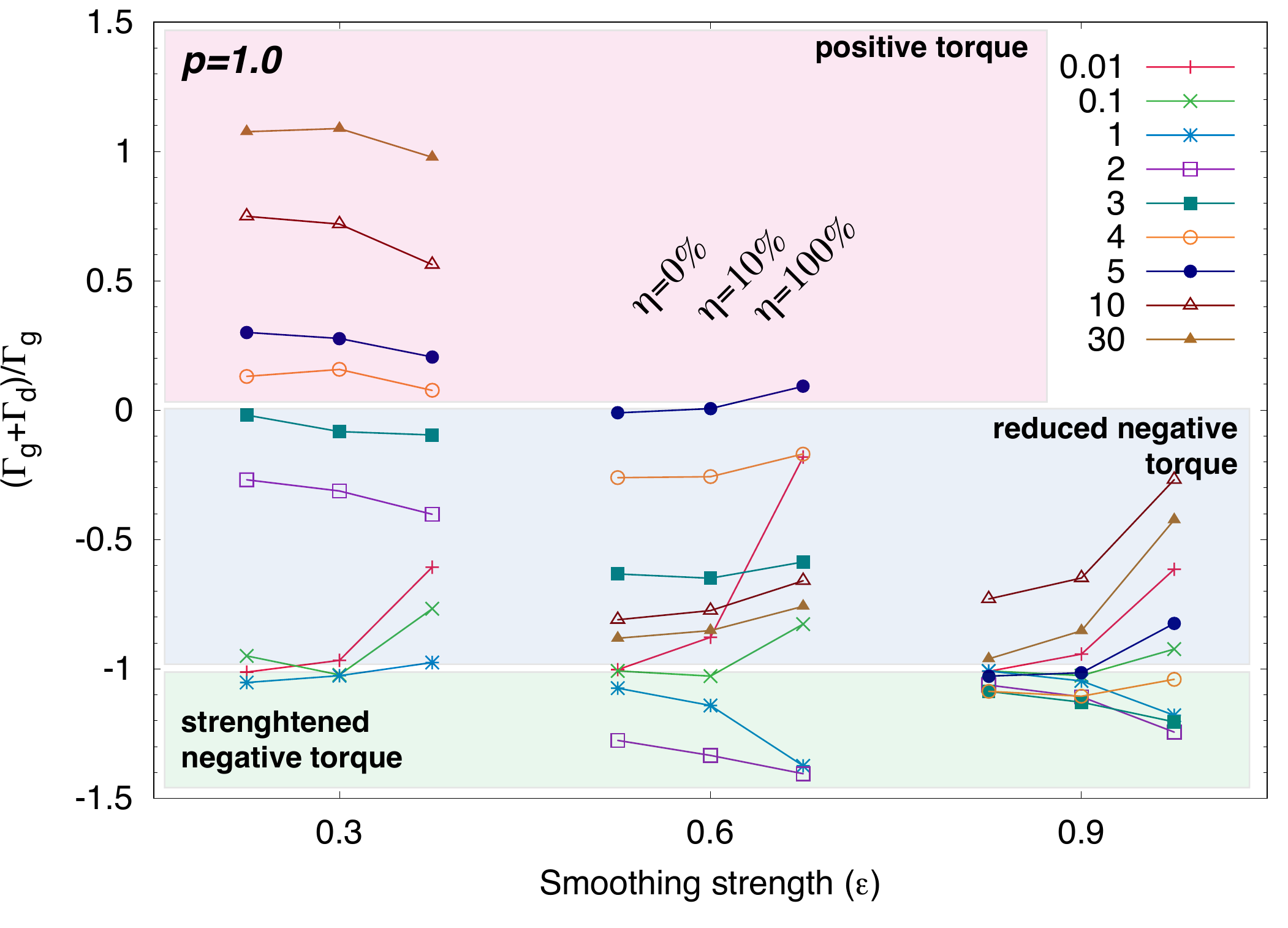}
   \caption{Effect of smoothing strength, $\varepsilon$, on the total torque normalised with absolute torque value of gas, $(\Gamma_\mathrm{g}+\Gamma_\mathrm{d})/|\Gamma_\mathrm{g}|$, felt by $1\,M_\oplus$ planet for different solid species. Result are shown for three different smoothing strength, $\varepsilon=0.3,\,0.6,\,0.9$ and accretion strength, $\eta=0,\,0.1,\,1$. The slope of initial gas density is set to $p=1.0$. Shaded regions are the same as in Figure~3.}
    \label{fig:trq-H}
\end{figure}

\begin{figure}
	\includegraphics[width=1\columnwidth]{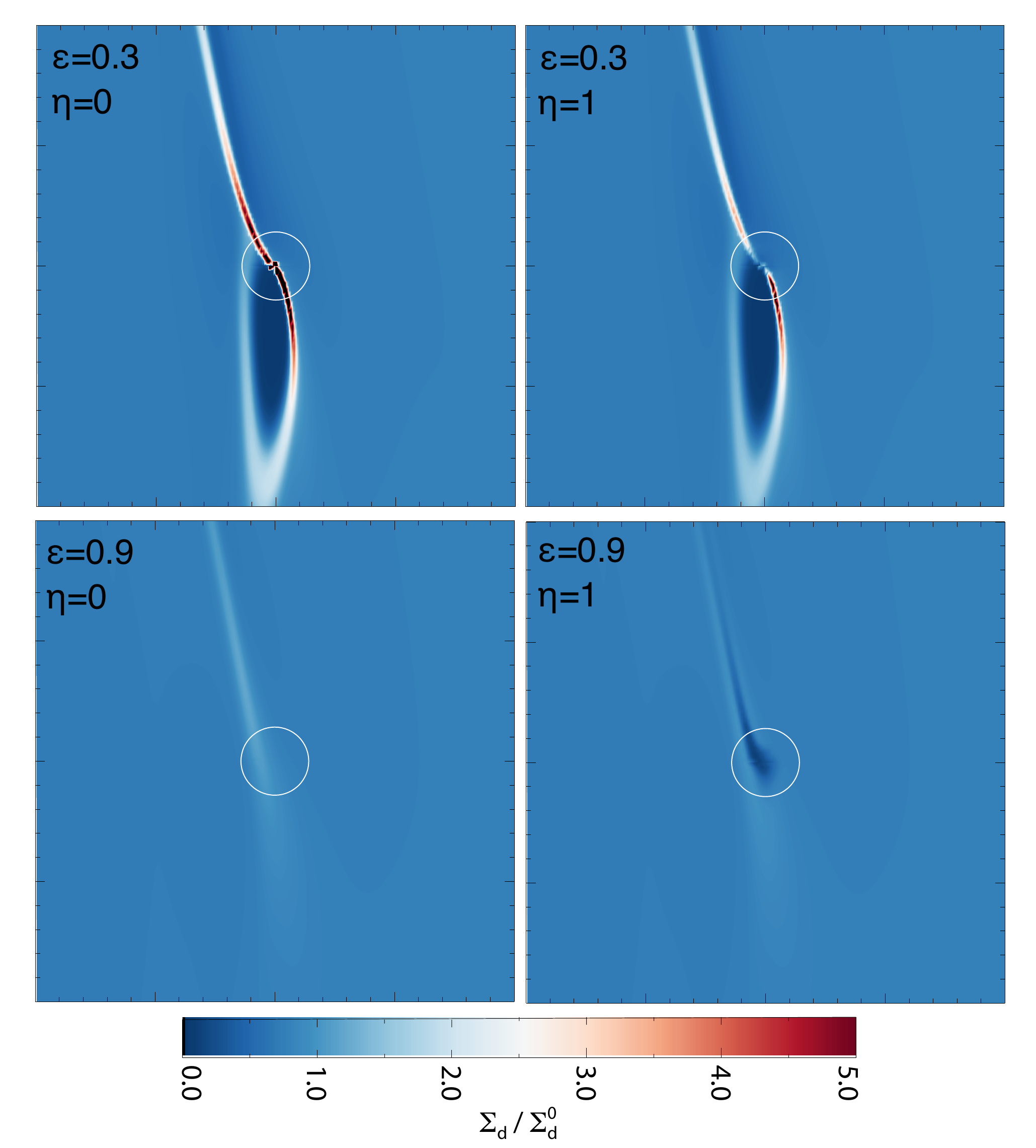}
    \caption{Comparison of the density distribution of $\mathrm{St}=1$ solid close to the planet assuming different smoothing strength of planetary potential $\varepsilon=0.3$ (upper panels) and $\varepsilon=0.9$ (lower panels). Non-accreting ($\eta=0$) and strong accreting ($\eta=1$) models are shown on left and right panels, respectively. The solid density, $\Sigma_\mathrm{d}$, is normalised with that of initial, $\Sigma_\mathrm{d}^0$.}
    \label{fig:comp-H}
\end{figure}

Due to numerical issues and for approximating three-dimensional effects it is necessary to smooth the planetary potential in two-dimensional simulations. In this study, we use a conventional method of potential smoothing with $\varepsilon=0.6$, see Equation~(\ref{eq:phi_tot}). In this method, the smoothing length, $\varepsilon H$, is proportional to the local pressure scale-height. However, solids are subject to sediment to the disk midplane. As a result, the vertical scale-height of solids differs from that of gas (see, e.g., \citealp{DullemondDominik2004}). In a simple approach, the solid scale height can be given as 
\begin{equation}
    H_\mathrm{d}=H \sqrt{\frac{\alpha}{St + \alpha}},
\end{equation}
in which case the well-coupled solid species have the same scale-height as the gas, while decoupled solid particles sink to the disk midplane  \citep{Birnstieletal2016}. This process is affected by the vertical turbulent mixing, which depends on the magnitude of viscosity via $\alpha$. The appropriate method of smoothing the planetary potential for solid species in two-dimensional simulations is unknown yet. However, the importance of smoothing strength can be revealed by changing the strength of smoothing. 

We investigated the effect of smoothing strength by applying $\varepsilon=0.3$ and $0.9$ for a $M_\mathrm{p}=1\,M_\oplus$ planet embedded in a disc having $p=1.0$ slope of initial density profiles in the three accretion regimes, see Fig.~\ref{fig:trq-H}. It is found that the gas torque depends on $\varepsilon$: $\Gamma_\mathrm{g}/\Gamma_0=1.82$ and $0.7$ for $\varepsilon=0.3$ and $0.9$, respectively. A similar trend can be observed for solid torques.  For the well-coupled solids ($\mathrm{St}=0.01$ and $0.1$). the change in smoothing strength has a week effect. However,  the magnitude of solid torques decreases with increasing $\varepsilon$ relative to the gas torque for $\mathrm{St}\geq1$.  Interestingly, the effect of accretion strength on solid torques becomes very strong in models that use $\varepsilon=0.9$.  For weak smoothing ($\varepsilon=0.3$), the effect of solids is so strong that even torque reversal can be observed for the less coupled solid species ( $\mathrm{St}\geq4$).

Fig.~\ref{fig:comp-H} compares the density distribution of $\mathrm{St}=1$ solid in the vicinity of $1~M_\oplus$ planet in non-accreting and strong accreting models assuming $\varepsilon=0.3$ and 0.9. It is appreciable that the dust depletion beyond the planet is prominent in $\varepsilon=0.3$ and completely missing in $\varepsilon=0.9$ model. Note that this solid depletion is also missing in our standard $\varepsilon=0.6$ models, see Fig.~\ref{fig:comparison} for comparison. The same can be observed for $\mathrm{St}\geq1$ species. Generally, the strength of solid asymmetry weakens with increasing strength of planetary potential smoothing. Therefore the total torques exerted by solids also decreasing with $\varepsilon$.

\section{Summary and Conclusion}
%----------------------------------------------------------------

We investigated the effect of solid accretion on the torque felt by a low-mass planet with the mass in the range $0.1M_\oplus-10 M_\oplus$, embedded in a protoplanetary disc using two-dimensional grid-based locally isothermal hydrodynamic simulations. We used $\alpha$-prescription for the disc viscosity in a low-viscosity regime $\alpha=10^{-4}$.  The disc self-gravity and solid back-reaction are neglected.  For simplicity, we modelled solids assuming constant Stokes numbers in the range $0.01-30$.  The gas-to-solid mass ratio assumed to be the canonical value of 0.01. We found that the accretion of solids  can be important with regards to the magnitude or even the sign of the torque felt by the low-mass planet. Our key findings are the followings:

(i) As a result of solid accretion, the spatial asymmetry developed in the solid distribution in the vicinity of the planet strengthens. The magnitude of asymmetry depends on the mass of the accreting planet, the Stokes number of solids, and the accretion strength. We found that the magnitude of solid torques (either being positive or negative) increases with accretion  strength.

(ii) The effect of solid accretion on the total torque is significant for low-mass, $M_\mathrm{p}\lesssim 1\,M_\oplus$, planets. The solid torque magnitude can overcome that of gas for vigorously accreting low-mass planets, which can cause either total torque reversal or strengthened negative total torque. Planets with $M_\mathrm{p}1\,M_\oplus$ generally experience reduced negative total torque independent of solid accretion strength.

(iii) The steepness of radial profiles of gas, $p$, affects the total torque felt by the planet. Planets generally feel more negative total torque in case of steeper profiles due to the larger magnitude of the negative gas torque.

(iv) Accretion of well-coupled solids ($\mathrm{St}\leq0.1$) can cause a very strong positive solid torque for a $M_\mathrm{p}<1\,M_\oplus$ planet independent of $p$. In contrast, accretion of larger solids ($\mathrm{St}\geq0.1$) generally causes increased magnitude negative total torque.  For an Earth-mass planet, accretion of $2\leq\mathrm{St}\leq5$ solid material causes large positive solid torque, which results in positive total torque in $p=0.5$ discs.  However, in $p\geq1.0$ discs, the accretion of same sized solid species results in weaker but negative total torque for an Earth-mass planet.  Solid torques are found to be insensitive to accretion strength for several Earth-mass, $M_\mathrm{p}\geq3\,M_\oplus$, planets.  As a result of positive solid torque, those planets feel weaker negative total torque compared to the analytical prediction of the canonical type-I approximation.

(v) Care must be taken in two-dimensional simulations as the effect of the smoothing of planetary potential can be significant on the gas and solid torques felt by the planet. We found that the weaker the smoothing (i.e., the smaller the value of $\varepsilon$) is, the stronger is the effect of solid on the planetary torque.

Let us consider a scenario, in which the accretion of solids is {\bf strong}, i.e., $\eta=1$ and the protoplanetary disc has a shallow initial radial profile ($p=0.5$). At the beginning of planet formation, it is plausible to assume that the growing planet is in the low-mass regime (e.g. $M_\mathrm{p}\lesssim0.1\,M_\oplus$), while the majority of solid material is well-coupled to the gas (i.e, $\mathrm{St}\simeq0.01$) at the planet-forming region (see Fig.\ref{fig:d-St}). As we have shown, a solid accreting low-mass planet feels positive total torque (see the upper panel of Fig.\ref{fig:gamma_end}). As a result, a small mass planet migrates outward, which continues as long as  $M_\mathrm{p}\simeq0.3\,M_\oplus$ and solids are not grown above $\mathrm{St}\simeq0.1$. Meanwhile, both the planet and solid material grow. Assuming that the planet has grown to Earth-mass and solids are in the pebble regime ($\mathrm{St}\simeq1$), the migration is still directed outward. By the time the planet has grown above several Earth-mass and solids size reaches the boulder regime ($2\leq\mathrm{St}\leq5$), its migration reverses. Emphasise that the rate of inward migration is under the prediction of the canonical type~I regime duo to the positive solid torque. By assuming a steeper initial radial density profile for the disc, $p\geq1.0$, the migration reversal can only occur below Earth-mass. 

An important effect of the solid accretion is that low-mass protoplanets ($<1~M_\oplus$) migrate faster if the solid material is in the pebble-sized regime. Thus, the survival of a solid accreting migrating protoplanet might be uncertain in an evolved disc, in which the majority of solids are in the pebble-sized regime. In other words, plant formation might be compromised in evolved discs due to solid accretion. Finally, we emphasise that we used a canonical value of for the solid-to-gas mass ratio $epsilon=0.01$. Since the solid torque magnitudes are linearly scaled with $\epsilon$, low-mass planets outward or the fast inward migration is slower in solid depleted discs. In contrast, if the solid content of the disc is above the canonical value (by only several times, e.g. $\epsilon=0.05$) the migration speed increases significantly.

The above-mentioned hypothetical migration history, however, can be altered by the disc model applied being two-, or three-dimensional and the details of solid accretion phenomenon. In this study, the solid accretion was prescribed in a simplistic approach. By analysing the flow pattern around a low-mass planet, \citet{Ormel2013} has shown that the planetary atmosphere is asymmetric and the accretion rate of well-coupled dust can be reduced. Thus, to better understand the effect of solid accretion on the torque felt by low-mass planets, it is worth reproducing our simulations by handling solid material as Lagrangian particles like in \citet{MorbidelliNesvorny2012}. 

Due to the necessity of high numerical resolution, our investigation has been done in two-dimensional thin disc approximation. Although some of the three-dimensional effects are taken into account (e.g., by applying gravity softening), the gas and dust flow around the planet is a three-dimensional phenomenon (see, e.g., \citealp{DAngeloetal2003,BitschKley2011,Legaetal2014,FungArtymowicz2015}). Moreover, as the appropriate magnitude of smoothing of the gravitational potential of the planet for solid species is unknown yet, it would be important to investigate the dynamics of solids  and accretion in three-dimension.

For simplicity, we applied a simple disc thermodynamics, i.e. locally isothermal approximation. Since the effect of negative entropy gradients in adiabatic disc or the accretion heating may result in outward migration (see, e.g., \citealp{PaardekooperMellema2006,Paardekooperetal2010,MassetCasoli2010,Benitez-Llambay2015}) it is worth study the effect of solid accretion in adiabatic discs too. 

Diffusion of solid species was neglected in this study. Therefore, we used nearly inviscid models, for which case it is plausible to assume that diffusion is negligible. Since diffusion may smear out the asymmetric distribution of solids, it is worth investigating the effect of viscosity and solid diffusion on the total torque felt by solid accreting planets.

Finally, we note that the inclusion of the solid feed-back might also be important. However, its effect can be minor as the maximum of the solid-to-gas mass ratio measured in the our simulations is about $\epsilon\simeq0.04$.

\section*{Acknowledgements}
This project was supported by the Hungarian OTKA Grant No. 119993 and by OeAD-OMAA program through project 95\"ou13. I gratefully acknowledge the support of NVIDIA Corporation with the donation of the Tesla 2075 and K40 GPUs. We thank for the usage of MTA Cloud (https://cloud.mta.hu/) which significantly helped us achieving the results published in this paper. Discussion on the numerical solution of solid dynamics with E. Vorobyov is acknowledged.  

%%%%%%%%%%%%%%%%%%%%%%%%%%%%%%%%%%%%%%%%%%%%%%%%%%



%\bibliographystyle{mnras}


%%%%%%%%%%%%%%%%%%%%%%%%%%%%%%%%%%%%%%%%%%%%%%%%%%

%%%%%%%%%%%%%%%%% APPENDICES %%%%%%%%%%%%%%%%%%%%%
\appendix

\section{Dust solver test}

In order to test the numerical solver for the solid species (see Equations\,(\ref{eq:solidsol-1})-(\ref{eq:solidsol-2}), the analytical solutions of velocity component of solids (see Equations\,(\ref{eq:vel3})-(\ref{eq:vel4})) are compared to the numerical solutions. In these simulations there is no embedded planet in the disc, i.e. the disc is unperturbed. Initially solid species have zero radial velocity and Keplerian azimuthal velocities.

Top and bottom panels on Figure\,\ref{fig:GFARGO2-test} compare the radial and azimuthal velocity (Keplerian velocity is subtracted) components resulted after 50 orbits of the planet and the analytical solution, respectively. One can see that the numerical solution perfectly fits the analytical one independent of the solid's Stokes number.

\begin{figure}
	\includegraphics[width=1\columnwidth]{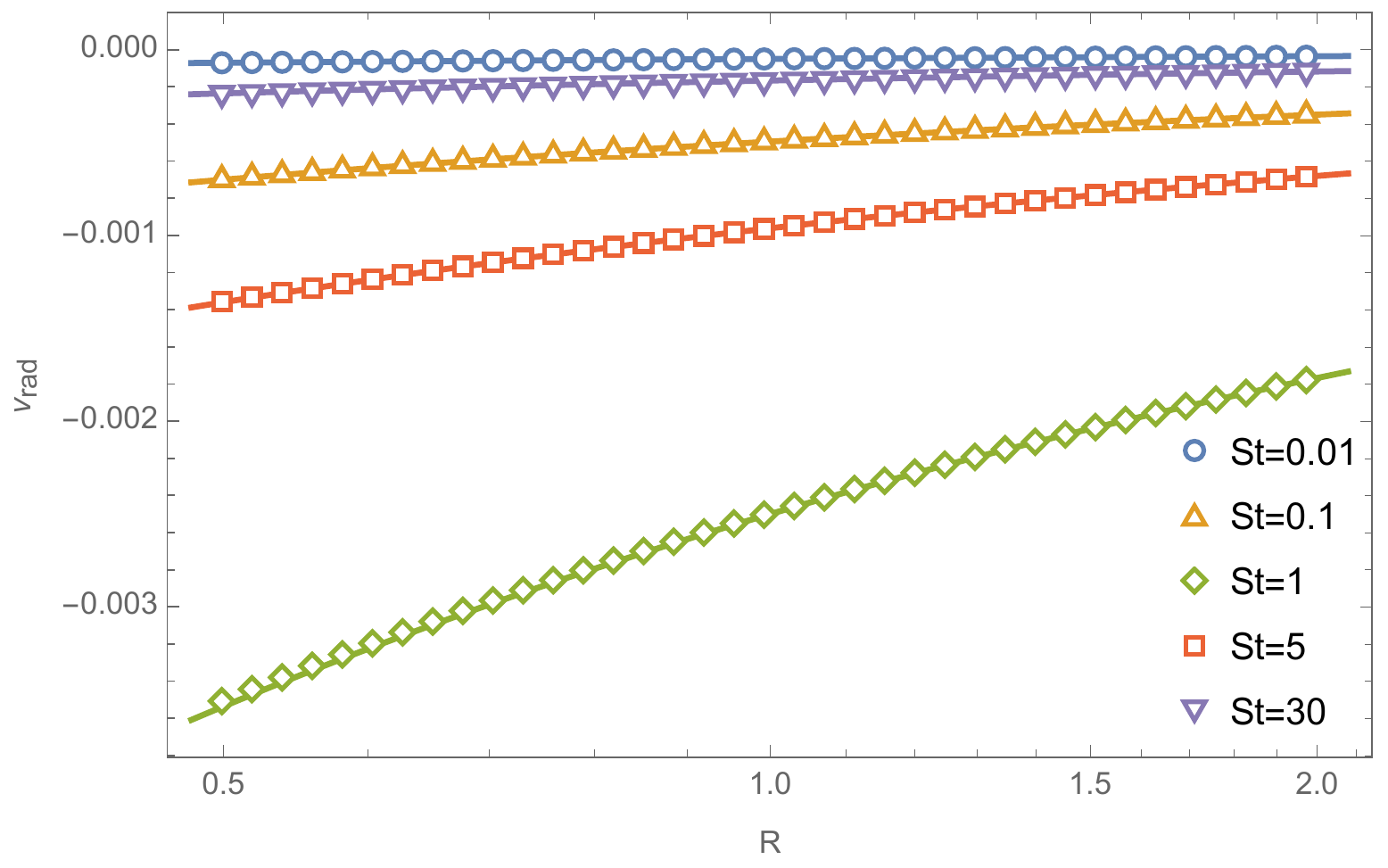}
	\includegraphics[width=1\columnwidth]{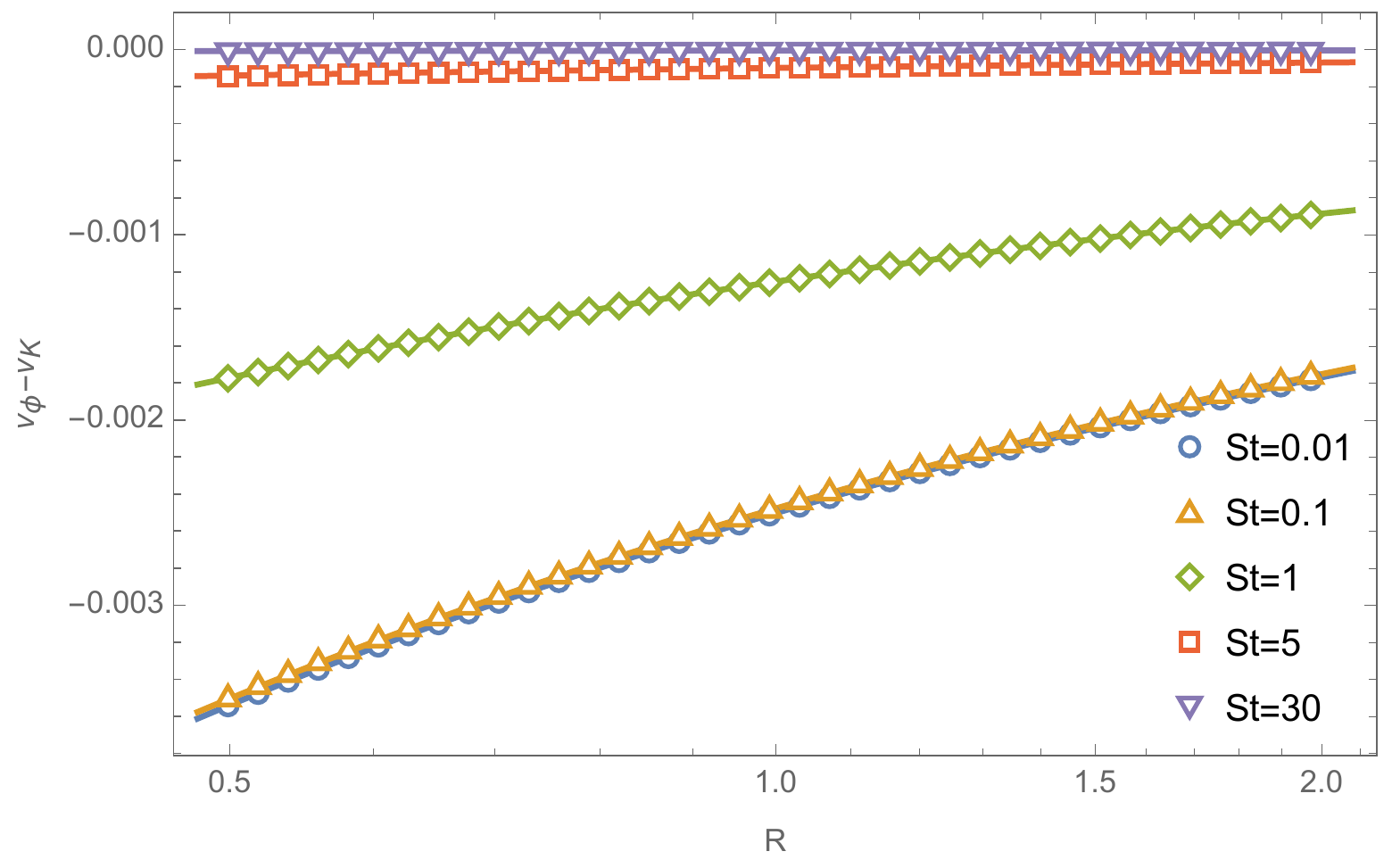}
   \caption{Comparison of the numerical to the analytical solution of solid dynamics. Radial velocity ($v_R$) and azimuthal minus Keplerian ($v_\phi-v_K$) components are shown in the upper and lower panel, respectively. Markers and solid lines represent the numerical and analytical solutions for the given Stokes number, respectively.}
    \label{fig:GFARGO2-test}
\end{figure}

%%%%%%%%%%%%%%%%%%%%%%%%%%%%%%%%%%%%%%%%%%%%%%%%%%

% Don't change these lines
%\bsp	% typesetting comment
\label{lastpage}
\end{document}